\documentstyle[twocolumn,prd,epsf,aps]{revtex}

\begin{document}                                                

\wideabs{

\draft

\title{From cool pions to the chiral phase transition}

\author{Marcello B. Silva Neto}

\bigskip

\address{
Instituto de F\protect\' \i sica, Universidade Federal do Rio de Janeiro,\\
Caixa Postal 68528, Rio de Janeiro - RJ, 21945-970, Brazil
}

\date{\today}
\maketitle


\begin{abstract}

Using the ideas of effective field theory and dimensional reduction, we relate the 
parameters of two low energy models of QCD: the $O(N)$ nonlinear sigma model in 
$D=3+1$, which describes the dynamics of cool pions, and the $O(N)$ Heisenberg magnet
in $D=3+0$, which is commonly argued to reproduce the correct critical behaviour of
the chiral phase transition. As a result, we obtain a generalized expression for the finite 
temperature pion decay constant which reproduces, in certain limits, the available expressions 
in the literature. 

\end{abstract} 

\pacs{PACS numbers: 11.10.Wx, 11.10.Ef, 11.10.Gh} 

} 

\begin{narrowtext}


The thermodynamic properties of the low energy dynamics of QCD can be apropriately 
described in terms of different $O(N)$ $\sigma$ models, deppending on whether or not 
we are close to the critical temperature $T_{c} \simeq 150$ Mev. At low temperatures, 
the only degrees of freedom which are excited are pions. One may then use the results 
of  chiral perturbation theory, which in its lowest order (very lowest energies) is 
just the nonlinear $\sigma$ model. In fact, as first pointed out by Weinberg 
\cite{Weinberg}, a suitable effective field theory involving Goldstone fields automatically 
generates transition amplitudes which obey the low energy theorems of current algebra 
and PCAC. The interaction among the Goldstone bosons is described
by an effective Lagrangian, which is invariant under global chiral transformations. 
Near the critical temperature, on the other 
hand, it has been demonstrated that the long wavelength fluctuation modes of the massless
$N_{f}=2$ QCD belong to the same universality class of the $O(4)$ Heisenberg magnet in 
three spatial dimensions \cite{Pisarski-Wilczek,Wilczek-Rajagopal}. The reasoning behind 
this proposal is based on counting the light degrees of freedom. The transition region 
is dominated by the longitudinal and transverse fluctuations of the order parameter, 
the $\sigma$ and $\vec{\pi}$ fields, which go soft at the transition temperature. Being bosonic, 
$\sigma$ and $\vec{\pi}$ have zero frequency Matsubara modes, $\omega_{n}=0$, which turn 
to be the only relevant degrees of freedom in the scaling region. The fermions themselves, 
even if they are massless at zero temperature, do not influence the nature of the phase 
transition at finite temperature. 

Before proceeding, however, we should mention that although this is a very intuitive scenario, 
there still is some divergence concerning the nature of the critical properties of massless 
$N_f=2$ QCD. For example, when one uses lattice simulation to analyse the critical behavior 
of the order parameter (the $q\bar{q}$ condensate), for which universality arguments suggest
\begin{equation}
\left< q\bar{q} \right> \sim
\left| (T-T_{c})/T_{c} \right|^{0.38 \pm 0.01},
\end{equation}
one concludes that lattice data is indeed consistent with the $O(4)$ spin system
\cite{Lattice-Data}. Conversely, when one tries to check universality arguments 
by looking at the behavior of global thermodynamical quantities, such as the 
specific heat, one finds that, while in the $O(4)$ spin system the singular 
contribution of the soft modes vanishes at the critical point, lattice data for 
the $N_{f}=2$ QCD actually show a huge peak in the specific heat around $T_{c}$.
As pointed out by Shuriak \cite{Shuriak}, this certainly mean that near 
criticality many new degrees of freedom become available or are significantly 
changed. This discrepancy may be understood if we remember that, while critical
correlation functions are dominated by the long wavelength fluctuation modes,
global thermodynamical quantities, such as the free energy or the specific heat, 
receive contribution from all energy modes. In fact, in any asymptotically free 
theory, like QCD, the effect of raising the temperature is to populate the weakly 
coupled higher energy modes \cite{Dine-Fischler}. In this sense, corrections to a 
free gas should fall with the temperature. The $n=0$ mode makes only a perturbatively 
small contribution and all other higher modes must be considered. The conclusion is 
that global thermodynamic quantities simply cannot be calculated using classical 
statistical mechanics.

In this brief report we will not enter into the above discussion but rather study further
the relation between the parameters of the two effective models for the low energy dynamics 
of QCD, from the point of view of effective field theory and dimensional reduction. This 
will be done with the aid of a ($3+1$)-dimensional $O(N)$ linear $\sigma$ model which will 
serve as a link between the above two effective models. Let us begin by considering the Lagrangian 
density
\begin{equation}
{\cal L}(\vec{\phi}(x))=
\frac{1}{2}(\partial_{\mu}\phi^{a})^{2}+
\frac{1}{2}\mu^{2}\phi^{a}\phi^{a}+
\frac{\lambda}{4}(\phi^{a}\phi^{a})^{2}, 
\label{Lagrangian-Density}
\end{equation}
which describes the long distance properties of a Heisenberg ferromagnet, where 
$\vec{\phi}(x)$ is a $N$-component real massive scalar field playing the role of 
a slowly varying order parameter. In the ordered phase ($\mu<\mu_{c}$), the field 
has a nonvanishing expectation value 
\begin{equation}
\left< \vec{\phi}(x)\cdot\vec{\phi}(x) \right>=M^{2}(\mu,\lambda)=\frac{-\mu^{2}}{\lambda},
\label{Expectation-Value}
\end{equation}
breaking spontaneously the $O(N)$ rotational symmetry to $O(N-1)$ invariance.
In order to consistently quantize the theory a prefered direction in the
internal space must be choosen. This choice is completely arbitrary, and
for our purposes it will be convenient to choose the direction of the 
unitary $N$-component field $\hat{\phi}(x)\equiv\vec{\phi}(x)/|\vec{\phi}(x)|$.
This can be implemented by the following {\it change of variables} in the 
functional integral
\begin{equation}
\vec{\phi}(x)=\rho(x)\hat{\phi}(x),
\label{Change-Variables}
\end{equation}
where $\hat{\phi}(x)\cdot\hat{\phi}(x)=1$ by construction.
In terms of the new variables $\rho(x)$ and $\hat{\phi}(x)$, the functional
integral becomes
\begin{equation}
Z=\int \rho^{N-1}(x){\cal D}\rho(x) {\cal D}\hat{\phi}(x)
e^{-S(\rho,\hat{\phi})},
\end{equation}
with
\begin{eqnarray}
S(\rho,\hat{\phi})&=&
\int {\rm d}^{4}x 
\left\{ 
\frac{1}{2}\rho^{2}(x) \left( \partial_{\mu}\hat{\phi^{a}}(x) \right)^{2} \right. \nonumber \\
&+&\left. \frac{1}{2} \left( \partial_{\mu}\rho(x) \right)^{2}+
\frac{1}{2} \mu^{2} \rho^{2}+
\frac{1}{4} \lambda \rho^{4}
\right\}.
\label{Action-Rho-Hat-Phi}
\end{eqnarray}
The theory has a natural UV cutoff, namely the meson mass $M$ defined by 
(\ref{Expectation-Value}). We can then perform integration over the $\rho(x)$ field 
in order to generate an effective local action $S_{eff}(\hat{\phi})$ for the 
long wavelength fluctuations of $\hat{\phi}$ 
\begin{equation}
e^{-S_{eff}(\hat{\phi})}=\int \rho^{N-1}(x){\cal D}\rho(x)
e^{-S(\rho,\hat{\phi})}.
\end{equation}
Although it is generally not possible to compute this integral exactly, due to the
$\rho$ quartic self-interaction, one can nevertheless compute it perturbatively.
Indeed, since $M^{2}(\mu,\lambda)=-\mu^{2}/{\lambda}$ is the vacuum 
expectation value of the field $\rho$, we can write
\begin{equation}
\rho(x)=M+\rho^{\prime}(x),
\end{equation}
where $\rho^{\prime}$ are fluctuations of the $\rho$ field around $M$. In terms of 
the new variable $\rho^{\prime}$ the action (\ref{Action-Rho-Hat-Phi}) reads
\begin{eqnarray}
S(\rho^{\prime},\hat{\phi})&=&
\int {\rm d}^{4}x 
\left\{ 
\frac{1}{2} \left[ M^{2} + 2M\rho^{\prime} + {\rho^{\prime}}^{2} \right] 
(\partial_{\mu}\hat{\phi^{a}}(x))^{2}
\right. \nonumber \\
&+&
\left.
\frac{1}{2} (\partial_{\mu}\rho^{\prime}(x))^{2}+
\frac{\mu}{2} \left[ M + \rho^{\prime} \right]^{2}+
\frac{\lambda}{4} \left[ M + \rho^{\prime} \right]^{4}
\right\} \nonumber.
\end{eqnarray}
The zeroth order in a perturbative expansion of the above effective field theory
is obtained by neglecting all $\rho^{\prime}$. As a result we obtain, after the
identification $M^{2}=f_{\pi}^{2}$, the expression
\begin{equation}
S_{eff}^{0}(\hat{\phi})=f_{\pi}^{2}\int {\rm d}^{4}x 
\frac{1}{2}\left[ \partial_{\mu}\hat{\phi^{a}}(x) \right]^{2},
\label{Eff-Action-Hat-Phi}
\end{equation}
which is simply the action of a nonlinear $\sigma$ model. It must be emphasized that 
loop corrections coming from the $\rho^{\prime}$ integration renormalize the coeficient 
$f_{\pi}^{2}(\mu,\lambda)$ in front of (\ref{Eff-Action-Hat-Phi}) \cite{Zinn-Justin-Book}. 
Additional $\hat{\phi}$ interactions can be expanded in local terms, provided we are 
exploring momenta much smaller than any mass scale in the problem. In this sense, we
conclude that the nonlinear $\sigma$ model completely describes the long distance properties 
of a Heisenberg ferromagnet at low temperature.  
 
Let us now turn to the other extreme. Near the critical temperature it is the long 
wavelength fluctuation modes of QCD which dominates the scaling behavior of correlation 
functions. Thus, roughly speaking, at least these correlation functions should be 
described by statistical mechanics. In terms of the discrete frequency sum of 
finite temperature quantum field theory, this means that one needs to retain only the 
zero-frequency components of the fields. Keeping only these zero-frequency modes yields 
a field theory in a lower dimension with temperature dependent renormalized coupling 
constants. The reduced theory is an effective theory for the zero modes of the 
original fields and can be explicitly obtained from the original theory by integrating 
out the nonzero Matsubara frequencies of all fields, bosons and/or fermions. 

Dimensional reduction of the $O(N)$ invariant linear $\sigma$ model was performed
in \cite{Marcello}, to one-loop order, using a modified minimal subtraction scheme 
at zero momenta and $T_{0}\neq 0$. It was shown that, from the Lagrangian density 
(\ref{Lagrangian-Density}), integration over nonstatic modes (nonzero Matsubara frequencies) 
give a dimensionally reduced effective free energy of the Landau-Ginzburg type for the 
bosonic zero mode $\phi_{0}$, $\vec{\phi}(x)=(\phi_{0}(x),0,...,0)$,
\begin{equation}
F(\phi_{0})=
\int {\rm d}^{3}{\bf x}
\left\{
          \frac{1}{2}(\partial_{i}\phi_{0})^{2}+ 
          \frac{1}{2}\mu_{R}^{2}\phi_{0}^{2}+
          \frac{\lambda_{R} T}{4}\phi_{0}^{4}
\right\},
\label{Reduced-Action}
\end{equation}
with all possible nonrenormalizable interaction terms supressed by the
temperature \cite{Landsman}. In the above expression, $\mu_{R}$ and $\lambda_{R}$ are the 
thermally renormalized mass and coupling constant given by \cite{Marcello}
\begin{equation}
\mu_{R}^{2}=\mu^{2}+(N+2)
\left\{
\frac{\lambda (T^{2}-T_{0}^{2})}{12}-\frac{\lambda \mu^{2}}{4\pi^{2}} 
\ln{\left( \frac{T}{T_{0}} \right)}
\right\}
\label{Renormalized-Mass}
\end{equation}
and
\begin{equation}
\lambda_{R}=\lambda-\frac{(N+8)}{6}\frac{\lambda^{2}}{16\pi^{2}}
\ln{\left( \frac{T}{T_{0}} \right)}.
\label{Renormalized-Coupling}
\end{equation}
From the above expressions it is clear that at long distances the only effect of the
nonstatic modes is to set the scale of the coupling constants to be the temeprature.
The dependence of both $\mu_{R}$ and $\lambda_{R}$ on the choice of the thermal 
renormalization point $T_{0}$ is controlled by a homogeneous renormalization group 
equation \cite{Marcello}
\begin{equation}
\left(
         T_{0}\frac{\partial}{\partial T_{0}}+
         \beta_{T_{0}}\frac{\partial}{\partial\lambda}+
         \gamma_{T_{0}}\mu^{2}\frac{\partial}{\partial \mu^{2}}
\right)
F(\phi_{0})=0,
\label{Thermal-RG-Equation}
\end{equation}
with renormalization group functions
\begin{equation}
\beta_{T_{0}}=\frac{(N+8)}{6}\frac{\lambda^{2}}{16 \pi^{2}}
\label{beta-function}
\end{equation}
and 
\begin{equation}
\gamma_{T_{0}}=(N+2)
           \left[ \frac{\lambda T_{0}^{2}}{12 \mu^{2}}+
                     \frac{\lambda}{16 \pi^{2}} \right],
\label{gama-function}
\end{equation}
in accordance to \cite{Landsman}.

Let us now define, in analogy to eq. (\ref{Expectation-Value}), the quantity
\begin{equation}
f_{\pi}^{2}(T,T_{0})=\frac{-\mu_{R}^{2}}{\lambda_{R}},
\label{Finite-T-Pion-Decay-Constant}
\end{equation}
which relates the parameters of the two low energy models of QCD, namely $f_{\pi}^{2}$ 
in (\ref{Eff-Action-Hat-Phi}) and $\mu_{R}$ and $\lambda_{R}$ in (\ref{Reduced-Action}).
It is not difficult to see that at $T_{0}=0$, and neglecting all the logarithms,
expression (\ref{Finite-T-Pion-Decay-Constant}) exactly reproduces the result of 
Bochkarev and Kapusta \cite{Bochkarev-Kapusta} for the finite temperature pion 
decay constant \footnote{There should be no reason to worry about the limit $T_{0}=0$ 
in expressions (\ref{Renormalized-Mass}) and (\ref{Renormalized-Coupling}). If we had 
adopted a renormalization prescription based on subtractions at $T_{0}=0$ the logarithms 
in expressions (\ref{Renormalized-Mass}) and (\ref{Renormalized-Coupling}) would be naturaly 
replaced by $\ln{(T/\sqrt{-\mu})}$ \cite{Marcello}.}
\begin{equation}
f_{\pi}^{2}(T,T_{0}=0)\equiv f_{\pi}^{2}(T)=
f_{\pi}^{2}\left[1-\frac{N+2}{12}\frac{T^{2}}{f_{\pi}^{2}}\right].
\end{equation}

We shall now give a physical interpretation to the quantity $f_{\pi}^{2}(T,T_{0})$. 
At zero temperature, $f_{\pi} \sim 93$ Mev and is connected to the matrix elements of the axial 
vector current ${\cal A}_{\mu}$ by $<0|{\cal A}_{\mu}^{a}|\pi^{b}(p)>=i f_{\pi} p_{\mu} \delta^{ab}$,
where the upper index in ${\cal A}_{\mu}^{a}$ is related to isospin. At finite temperatures, on 
the other hand, it has been recently proposed by Bochkarev and Kapusta \cite{Bochkarev-Kapusta},
using standard linear response theory, that the quantity $f_{\pi}^{2}(T)$ would measure the 
strength of the coupling of the Goldstone bosons to the longitudinal part of the axial spectral
density for ${\cal A}_{\mu}$, in the limit of zero external momentum. Alternatively, it was further
suggested in \cite{Jeon-Kapusta} that $f_{\pi}^{2}(T,0)/f_{\pi}^{2}$ would give the percentual of
(or the probability of finding) Goldstone excitations in the ground state. For this reason, it is 
possible to speak about a finite temperature phase transition in the system described by 
(\ref{Eff-Action-Hat-Phi}), associated to the complete decoupling of the Goldstone bosons from 
the ground state, which happens at a critical temperature \cite{Bochkarev-Kapusta}
\begin{equation}
T_{c}^{2}=\frac{12}{N+2}f_{\pi}^{2}.
\label{Critical-Temperature}
\end{equation}

Following the above discussion, we are then lead to give similar interpretation for the quantity 
$f_{\pi}^{2}(T,T_{0})/f_{\pi}^{2}$ as measuring the percentual of (or the probability of finding)
Goldstone excitations between the ground state and the state corresponding to $T_{0}$. This is in 
fact evident from (\ref{Renormalized-Mass}) and (\ref{Renormalized-Coupling}) from which we obtain 
$f_{\pi}^{2}(T,T_{0}=T)=f_{\pi}^{2}$, $\forall {\;} T$. This trivially states that the percentual 
of Goldstone excitations between the ground state and the state corresponding to $T_{0}=T$ is unity 
(or $100 \%$), as expected. In this sense, we conclude that expression (\ref{Finite-T-Pion-Decay-Constant}) 
generalizes the results of \cite{Bochkarev-Kapusta} by allowing us to have access to the dynamics 
of the Goldstone system in a wide energy region, ranging from the ground state $T_{0}=0 $ to the 
state corresponding to $T_{0}=T$.

Expression (\ref{Finite-T-Pion-Decay-Constant}) also reveals the existence of a critical line of second 
order phase transitions defined by
\begin{equation}
f_{\pi}^{2}(T_{c},T_{0})=0.
\label{Critical-Temperature-Line}
\end{equation}
In fact, by solving the above equation for $T_{c}$ we obtain, neglecting all the logarigthms,
\begin{equation}
T_{c}^{2}(T_{0})=\frac{12}{N+2}f_{\pi}^{2}+T_{0}^{2},
\end{equation}
which is consistent with the result (\ref{Critical-Temperature}) for $T_{0}=0$.
That $T_{c}^{2}(T_{0})$ is a monotonic growing function of $T_{0}$ should not be surprising.
It simply means that it is more and more likely to find excitations populating 
higher energy states, than in the ground state. 

We have obtained an expression for the finite temperature pion decay constant which 
somehow generalizes the results of \cite{Bochkarev-Kapusta}. We believe that our
expression (\ref{Finite-T-Pion-Decay-Constant}) may be useful in studies in which
one is interested in computing transition amplitudes between arbitrary energy states, 
other than the ground state. In fact, since the pion contribution to correlation functions
must depend on the probability of the system to be in a given energy state, the quantity
(\ref{Finite-T-Pion-Decay-Constant}) will act as a weight in computing correlation functions. 

The author is indebted with A. P. C. Malbouisson and N. F. Svaiter for
inumerous discussions and comments in early stages of this work. 
FAPERJ is also acknowledged for finantial support.

\end{narrowtext}

\end{document}